\documentclass[a4paper, fleqn, review]{cas-sc}

\usepackage[numbers, sort&compress]{natbib}

\usepackage{enumitem}
\graphicspath{{./figs}}
\usepackage{subcaption}
\usepackage{rotating}
\usepackage{float}
\usepackage{caption}
\usepackage[nameinlink, noabbrev]{cleveref}

\begin{document}
\let\WriteBookmarks\relax
\def\floatpagepagefraction{1}
\def\textpagefraction{.001}

\shorttitle{Rock Classification through Knowledge-Enhanced Deep Learning: A Hybrid Mineral-Based Approach}    

\shortauthors{Iye Szin Ang et al.}  

\title [mode = title]{Rock Classification through Knowledge-Enhanced Deep Learning: A Hybrid Mineral-Based Approach}



%

\author[1]{Iye Szin Ang}

\cormark[1]


\ead{iye.ang@unileoben.ac.at}


\credit{Conceptualization, Methodology, Software, Formal analysis, Investigation, Data Curation, Writing – Original Draft, Writing – Review \& Editing, Visualization}

\affiliation[1]{organization={Chair of Cyber-Physical-Systems, Montanuniversität Leoben},
            addressline={Franz Josef-Straße 18},
            city={},
            citysep={8700 Leoben}, 
            postcode={}, 
            country={Austria}}


\author[2]{Martin Johannes Findl}

\ead{martin.findl@unileoben.ac.at}

\credit{Methodology, Data Curation, Resources, Validation, Writing – review \& editing}

\affiliation[2]{organization={Chair of Waste Processing Technology and Waste Management, Montanuniversität Leoben},
            addressline={Franz Josef-Straße 18},
            city={},
            citysep={8700 Leoben}, 
            postcode={}, 
            country={Austria}}

\author[3]{Elisabeth Hauzinger}

\credit{Project administration, Resources}

\affiliation[3]{organization={Chair of Subsurface Engineering, Montanuniversität Leoben},
            addressline={Franz Josef-Straße 18},
            city={},
            citysep={8700 Leoben}, 
            postcode={}, 
            country={Austria}}


\author[2]{Klaus Philipp Sedlazeck}

\credit{Writing – review \& editing, Validation, Supervision}


\author[5]{Jyrki Savolainen}

\credit{Writing – review \& editing, Validation, Supervision}

\affiliation[4]{organization={LUT-kauppakorkeakoulu},
            addressline={Yliopistonkatu 34},
            city={},
            citysep={53850 Lappeenranta}, 
            postcode={}, 
            country={Finland}}


\author[5]{Ronald Bakker}

\credit{Validation, Supervision}

\affiliation[5]{organization={Chair of Resource Mineralogy, Montanuniversität Leoben},
            addressline={Franz Josef-Straße 18},
            city={},
            citysep={8700 Leoben}, 
            postcode={}, 
            country={Austria}}


\author[3]{Robert Galler}

\credit{Supervision, Funding acquisition, Project administration}


\author[1]{Elmar Rueckert}

\credit{Validation, Writing – review \& editing, Supervision, Project administration}


\begin{abstract}
Automated rock classification from mineral composition presents a significant challenge in geological applications, with critical implications for material recycling, resource management, and industrial processing. While existing methods using One-dimensional Convolutional Neural Network (1D-CNN) excel at mineral identification through Raman spectroscopy, the crucial step of determining rock types from mineral assemblages remains unsolved, particularly because the same minerals can form different rock types depending on their proportions and formation conditions. This study presents a novel knowledge-enhanced deep learning approach that integrates geological domain expertise with spectral analysis. The performance of five machine learning methods were evaluated out of which the 1D-CNN and its uncertainty-aware variant demonstrated excellent mineral classification performance (98.37±0.006\% and 97.75±0.010\% respectively). The integrated system's evaluation on rock samples revealed variable performance across lithologies, with optimal results for limestone classification but reduced accuracy for rocks sharing similar mineral assemblages. These findings not only show critical challenges in automated geological classification systems but also provide a methodological framework for advancing material characterization and sorting technologies.
\end{abstract}


\begin{highlights}
\item Raman spectroscopy enables automated mineral identification.
\item Weighted mineral composition system determines rock types.
\item Machine learning framework incorporates geological expertise.
\item Knowledge-infused ML pipeline facilitates automated mineral-to-rock classification.
\end{highlights}



\maketitle

\section{Introduction}\label{intro}
Rock classification and characterization are fundamental processes in the construction and mining industries, particularly for material recycling and sustainable resource management \cite{haasWasteValuableResource2020, haasApplicabilityExcavatedRock2021}. Accurate identification and analysis of rock types facilitate optimized resource utilization, enhance operational efficiency, and contribute to environmentally responsible practices. Traditional rock classification is primarily based on expert petrographic analysis, which integrates macroscopic observations, microscopic examinations, and manual mineral identification, often supplemented by chemical and/or mineralogical compositional data. However, automated approaches based on mineral composition present unique opportunities in applications that require automated, rapid and accurate classification.

This study establishes a systematic framework for automated rock classification that focuses on three representative rock types that are both geologically significant and economically important: granite, sandstone, and limestone. These rocks were selected based on three key criteria: (1) their widespread occurrence in the earth's crust, (2) their economic significance in the construction, energy, and mineral industries, and (3) their distinct mineralogical compositions that make them ideal candidates for automated classification systems.

Although existing mineral identification methods achieve high accuracy (>96\%) using Raman spectroscopy, there is a fundamental methodological gap in automated classification of rock types from these mineral assemblages. Consequently, the crucial step of automatically determining the rock types of these mineral assemblages remains an unresolved challenge. This study addresses the question: How can an automated system to accurately classify various rock types based on mineral assemblages identified through Raman spectroscopy be developed? To the best of our knowledge, this work presents the first systematic approach to automatically classify rock types directly from Raman-identified mineral assemblages.

A mineral-to-rock deduction classification system using Raman spectroscopic data was developed and evaluated to address this classification challenge. Raman spectroscopy was selected for its non-destructive analytical capabilities and its ability to provide distinct molecular fingerprints for different minerals, making it particularly suitable for mineral-based rock classification. The system was initially validated with granite before being expanded to sandstone and limestone. These rocks present varying mineralogical complexities, allowing for a systematic evaluation of the classification approach. By starting with this controlled scope, the potential for automated classification can be systematically assessed before expanding to more complex scenarios.

This approach combines a One-dimensional Convolutional Neural Network (1D-CNN) with domain-expert rules in a baseline system, leveraging both data-driven learning and established geological knowledge. An uncertainty-aware variant (1D-CNN-UNK) was also explored, designed to handle ambiguous cases and potential unknown mineral assemblages. Through this focused investigation, foundational insights for developing more comprehensive systems capable of handling multiple rock types in automated settings are aimed to be established.

The primary contributions of this research are summarized as follows:
\begin{enumerate}

    \item An integrated framework that combines a One-dimensional Convolutional Neural Network (1D-CNN) with a knowledge-enhanced expert system is proposed. This hybrid approach enables automated mineral identification while incorporating domain expertise through systematic knowledge integration. Using both data-driven learning and established geological knowledge, the framework addresses the limitations of traditional expert systems, which often rely solely on predefined rules or heuristics.

    \item A quantitative methodology for rock type classification through a percentage-based mineral composition weighting system is developed. This system accounts for the variations of mineral assemblages and their relative abundances in different geological settings, providing a more robust classification framework compared to traditional binary classification approaches. The methodology enhances classification accuracy by considering the nuanced compositional differences that are often overlooked in simpler models.

    \item The framework is validated using a carefully curated dataset comprising mineral spectra from the RRUFF database, structured according to expert-designed rock composition templates. This validation methodology ensures geological validity through systematic sampling of diagnostic mineral assemblages, expert-verified compositional relationships, and high-quality spectral data from standardized sources. The dataset's design and the validation process are described in detail to underscore the rigor and reliability of the findings.
\end{enumerate}

\section{Related Work}\label{work}
The application of Raman spectroscopy for rock and mineral identification has undergone significant advancements since its early laser-based implementations \cite{mcmillanRamanSpectroscopyMineralogy1989, clarkSpectroscopyRocksMinerals1999}. The integration of artificial intelligence (AI) has transformed spectral data processing and analysis by enabling the use of advanced machine learning algorithms that can handle large volumes of spectral data, leading to improved pattern recognition and classification capabilities. This transformation has resulted in more data being processed efficiently, better image processing techniques, and the development of comprehensive spectral databases. For example, AI-driven Raman spectroscopy has been successfully applied in the automated identification of minerals in geological samples, significantly enhancing the efficiency and accuracy of mineralogical studies \cite{julve-gonzalez.2023}.

The development of spectral databases has been crucial in advancing mineral identification through Raman spectroscopy, as it has enabled the creation of standardized reference spectra for thousands of minerals. The RRUFF project \cite{lafuente1PowerDatabases2015} can be regarded as one of the cornerstones of this domain, providing quality-controlled data, detailed crystallographic information, and documentation of sample origins and conditions. This standardized repository has been used for various applications, from portable gemstone identification systems \cite{culkaIdentificationGemstonesUsing2019} to machine learning and deep learning approaches \cite{careyMachineLearningTools2015, Liu_2017}. Recent developments have further expanded its utility through high-throughput computational methods \cite{bagheriHighthroughputComputationRaman2023} and open-source analysis tools \cite{Georgiev_2023}.

Progress in machine learning has transformed the analysis of Raman spectrum data. Qi et al. \cite{qiRecentProgressesMachine2023} provide a comprehensive review of these advances, documenting the progression from traditional statistical methods to more sophisticated deep learning approaches. Among these methods, 1D-CNN have emerged as particularly effective architectures for spectral data analysis \cite{acquarelliConvolutionalNeuralNetworks2017}, demonstrating excellent performance across various spectroscopic applications including chemical substance identification, molecular structure analysis, and material characterization. 

The practical application of automated Raman analysis extends to various industrial settings, particularly in sustainable resource management. Resch et al.~\cite{reschTunnelExcavationMaterial2009} demonstrated in their study of tunnel excavation materials that the proper characterization and classification of excavated materials could allow their recycling as high-value raw materials. Their work emphasizes not only the need for rapid material characterization and reliable identification methods but also the potential for automated systems to support sustainable construction practices through improved material recycling. 

Despite these advances, existing research has focused primarily on mineral-level identification through Raman spectroscopy and machine learning. Although these methods excel at identifying individual minerals, they seem to face fundamental limitations when applied to rock type classification. This is due to the fact that rocks are composite materials formed by varying combinations and proportions of minerals: the same minerals can occur in different proportions to form entirely distinct rock types. For example, both granite and sandstone contain mainly quartz and feldspar, but a granite is an igneous rock formed by magma crystallization, while a sandstone is a sedimentary rock formed by the compaction of mineral grains. Their different formation processes result in distinct textures and structures that define them as separate rock types, even though they share common minerals. Current research trends focus on enhancing mineral identification through improved data preprocessing and validation methodologies, yet there remains a critical gap in the literature: the lack of automated systems that can deduce rock types from identified mineral assemblages.

The remainder of this paper is structured as follows: \Cref{method} describes the methodology, including the development of the integrated framework and the quantitative methodology for rock type classification
\Cref{results} presents the results of the validation experiments and discusses the implications of the findings. Finally, \Cref{conclusion} concludes the paper and suggests future research directions.

\section{Metholodogy}\label{method}
A rock is a naturally occurring solid aggregate composed of minerals, fragments of minerals or rocks, remnants of organisms, or, in some cases, non-mineral substances. They typically comprise one or more rock-forming minerals and develop as a result of their specific geological setting. This study presents a systematic methodology for mineral assemblage-based rock classification using Raman spectroscopy, focusing specifically on the identification of three different rock types: granite, sandstone, and limestone. Our approach integrates 1D-CNN with knowledge-guided systems to create a robust classification framework. The methodology encompasses four key components: (1) a classification framework that establishes the theoretical foundation for mineral-based rock type identification, (2) systematic dataset collection and generation procedures, (3) development and implementation of two distinct mineral classification models: a baseline 1D-CNN approach and an uncertainty-aware model; and (4) a hierarchical confidence-based knowledge-guided system that take advantage of established geological knowledge for final classification decisions.

\begin{figure}
    \centering
    \includegraphics[width=\linewidth]{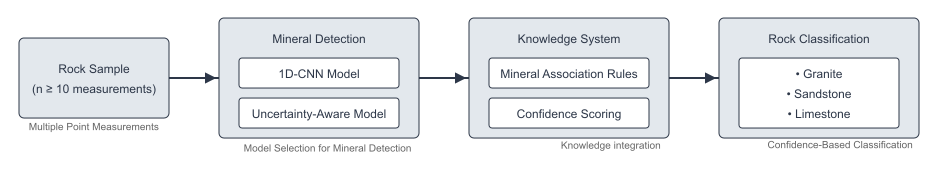}
    \caption{Schematic overview of the mineral-based rock classification framework incorporating knowledge-enhanced deep learning.}
    \label{fig:overview}
\end{figure}

Our methodology specifically targets the mineral-to-rock classification pipeline, rather than the broader conveyor belt implementation. The schematic overview in \Cref{fig:overview} demonstrates the knowledge-enhanced rock classification system. The workflow processes multiple measurement points (n $\geq$ 10) from each rock sample through either a standard 1D-CNN or its uncertainty-aware variant. The mineral detection output interfaces with a knowledge system that implemented expert-defined association rules and confidence scoring (parameterized by thresholds $\delta$c and $\delta$d). The final classification determines rock type (granite, sandstone, or limestone) based on both detected mineral assemblages and confidence metrics. This integrated approach combines data-driven learning with domain expertise to address the fundamental challenges in automated rock type classification from Raman spectroscopy measurements.

\subsection{System Assumptions and Theoretical Foundations}
The hybrid rock classification system relies on several key assumptions:
\begin{enumerate}[label=(\roman*)]
    \item It is assumed that rock classification can be effectively automated through the integration of Raman spectroscopy data and expert system rules;
    \item Each Raman measurement point sufficiently captures representative mineral assemblages of the rock sample;
    \item The presence and relative abundances of key mineral assemblages are sufficient for preliminary rock type determination;
    \item Expert geological knowledge can be effectively encoded into a rule-based system, including both standardized classification schemes and expert experience;
    \item Natural variations in mineral compositions and assemblages can be accommodated within defined confidence intervals;
\end{enumerate}

The classification rules in this system are derived from the geological literature. For igneous rocks (plutonic and volcanic), the classification framework is based on the QAPF (Quartz, Alkali feldspar, Plagioclase, Feldspathoid) diagrams established by the International Union of Geological Sciences (IUGS) \cite{streckeisen1976each, le1991iugs}. Here, the modal mineral content for the e.g. plutonic rocks (i.e. granite) is determined in relation to their quartz, alkali feldspar and plagioclase contents and normalized to 100 \% which places the rock in the corresponding field in the QAPF diagram, i.e. the granite field. Other rock forming mineral, e.g. micas, are not considered in this classification and are consequently neglected. This standardized classification scheme provides the theoretical foundation for the decision trees implemented in our expert system, particularly for granite and other plutonic igneous rock classifications.

For sedimentary rocks, such as sandstone and limestone, the classification rules were developed through knowledge elicitation from expert geologists. Sandstone classification is based on the content of the different grains, which are normalized to $100 \%$ and plotted in triangular diagrams with the corner points of the quartz - feldspar - lithic rock fragments. Two diagrams are used based on the fine-grained matrix (grain sizes $< 0.03$ mm) content. Over the years, numerous classification schemes for clastic sediments have been proposed. For sandstones with a grain size of $< 2$ mm, the scheme originally developed by Krynine (1948)~\cite{krynine1948megascopic} and later refined by Dott (1964)~\cite{dott1964wacke} and Pettijohn et al. (1987)~\cite{Pettijohn1987}  has become widely accepted (Okrusch \& Frimmel, 2022)~\cite{okruschMineralogieEinfuehrungSpezielle2022}. This ternary classification diagram is based on the relative proportions of Q (quartz), F (feldspar), and L (lithic fragments), which are normalized to a total of $100 \%$. Furthermore, variations in matrix content, typically characterized by mean grain sizes of $< 30 \mu$m, are represented along an axis perpendicular to the ternary diagram.  Rocks which contain $< 15 \%$ matrix are classified as arenites, $\geq  15 \%$ matrix as wacke $\geq  75 \%$ matrix are classified as claystone. We concentrate on our expert developed sandstone (arenite) decision tree on quartzitic/quartz sandstones with a matrix content of $0-15 \%$. Hence, the sandstone decision tree represents quartzarenit, sublitharenit, and subarkose~\cite{dott1964wacke, folk1968petrology, folk1974petrology}. The typical classification of limestones is based on the calcite-dolomite ratio. With respect to limestone, we concentrate on a typical limestone ($< 10\%$ dolomite) and a dolomitic limestone ($10-50 \%$ dolomite)~\cite{BGS2020, okruschMineralogieEinfuehrungSpezielle2022}. This process incorporated standard sedimentary rock classification schemes, expert field identification practices, practical experience in distinguishing key mineral assemblages, and common textural and compositional indicators.

For our mineral-based classification approach, the sandstone classification rules mentioned above were augmented by the presence of characteristic accessory minerals and the relative abundance of matrix materials. Similarly, the limestone classification incorporates carbonate mineral assemblages and common impurities. For granite, no minor compounds are considered in addition to the main minerals.

\subsection{Rock Classification Framework}
The automated classification of rocks from spectral data presents unique computational challenges due to the complex relationship between mineral assemblages and lithological classification. The proposed framework addresses these challenges through an integrated approach that combines spectral analysis with established petrological principles, implemented via a dual-layer classification architecture.

\subsubsection{Classification Overview}
\begin{figure}
    \centering
    \includegraphics[width=\linewidth]{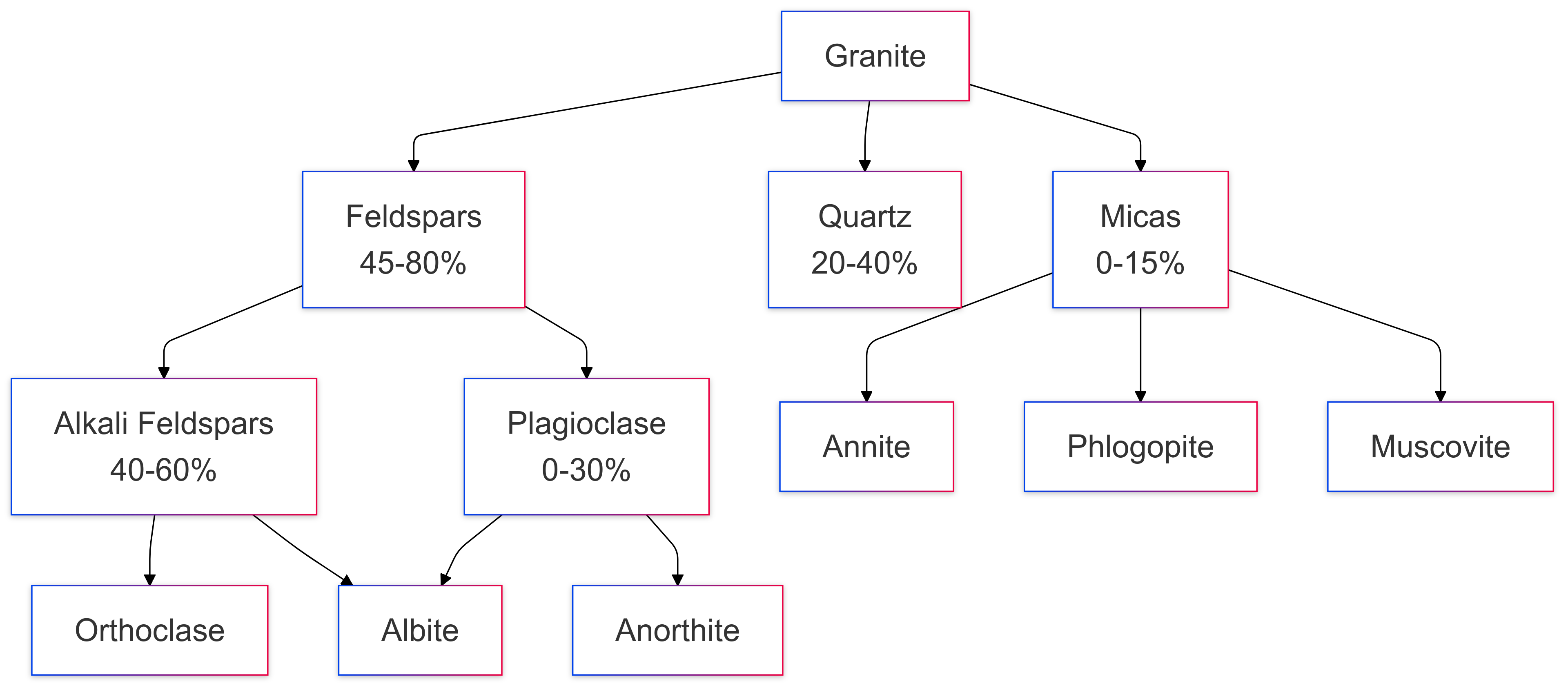}
    \caption{Hierarchical relationship between Granite and its essential minerals}
    \label{fig:granite-tree}
\end{figure}

This framework covers three rock types: granite, sandstone and limestone, which are formed through different geological processes under distinct chemical and physical properties, leading to varying mineral compositions. Granite, as an igneous rock, is primarily defined by its magmatic origin and characteristic mineral composition. The principal rock forming minerals include feldspars  (Albite, Anorthite, Orthoclase), mica group minerals (Annite, Muscovite, Phlogopite), and Quartz. \Cref{fig:granite-tree} shows the hierarchical relationship of granite-forming minerals and their typical proportions: feldspars ($45-80 \%$), quartz ($20-40 \%$), and mica minerals ($0-15 \%$) \cite{streckeisen1976each}. 

\begin{figure}
    \centering
    \includegraphics[width=\linewidth]{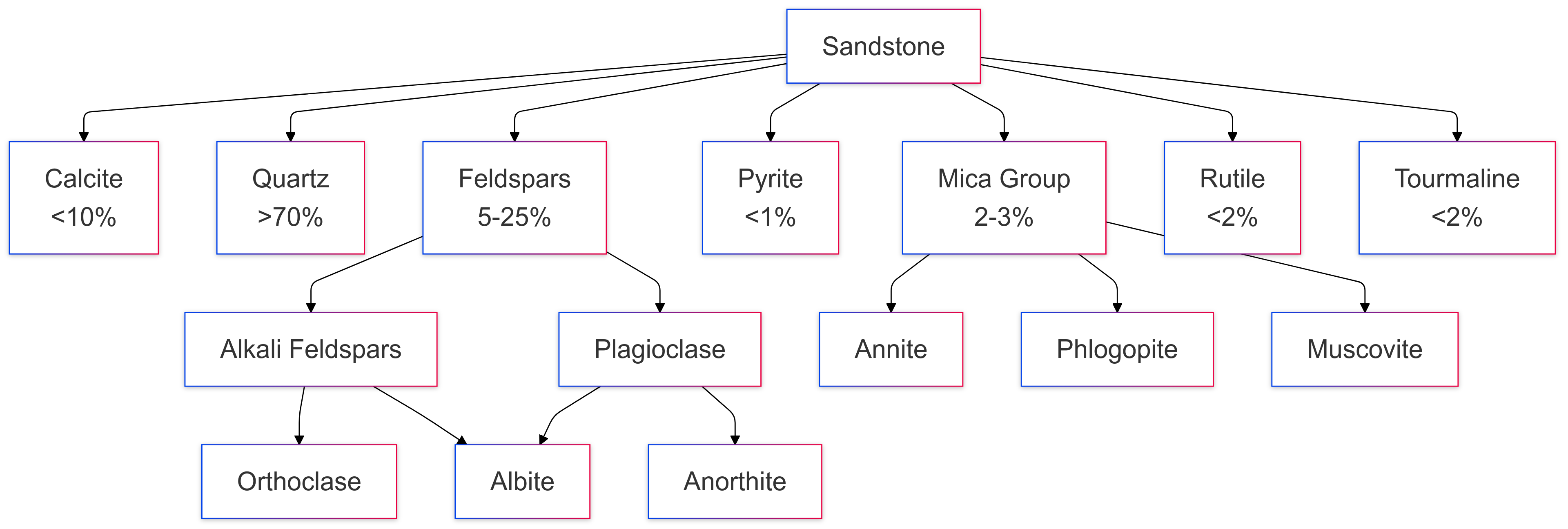}
    \caption{Hierarchical relationship between Sandstone and its essential minerals}
    \label{fig:sandstone-tree}
\end{figure}

Sandstone is a clastic sedimentary rock (\Cref{fig:sandstone-tree}) that is dominated by quartz ($> 70 \%$) with significant contributions from feldspars ($5-25 \%$, including both alkali feldspars and plagioclase) and minor components of calcite ($<10 \%$), pyrite ($< 1 \%$), mica group minerals ($2-3 \%$), rutile ($< 2\%$), and tourmaline ($< 2\%$). Calcite, pyrite, mica, rutile, and tourmaline were added with a realistic appearance to come as close as possible to the reality of a sandstone.

\begin{figure}
    \centering
    \includegraphics[width=.5\linewidth]{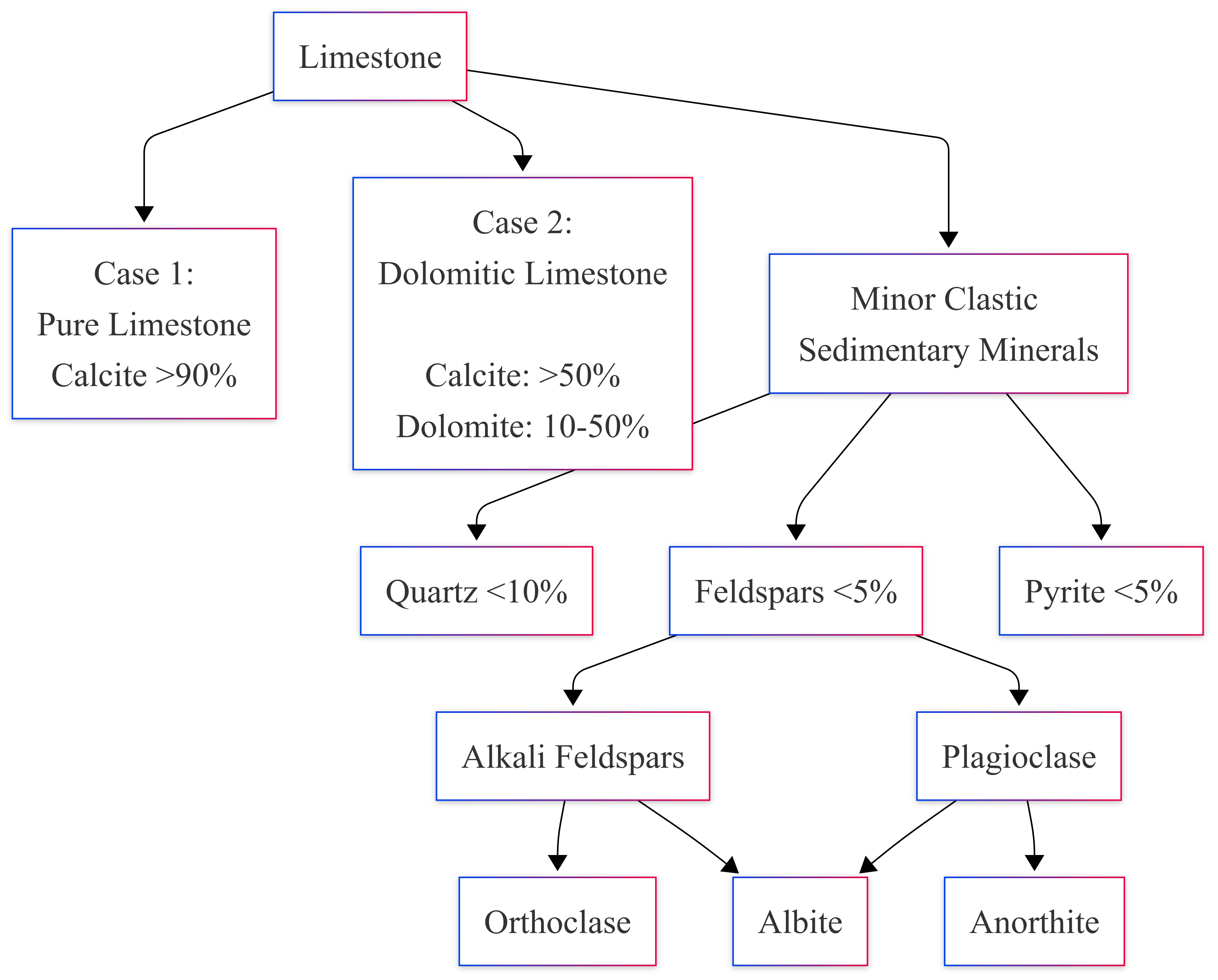}
    \caption{Hierarchical relationship between Limestone and its essential minerals}
    \label{fig:limestone-tree}
\end{figure}

The majority of carbonate rocks have been deposited in relatively shallow water in tectonically stable marginal areas. Carbonates have been formed by chemical or biochemical processes and also to a lesser extent by clastic sedimentation. These formation processes justify taking into account secondary and accessory minerals such as quartz \cite{Pettijohn1987, flugelMicrofaciesCarbonateRocks2004}. As illustrated in \Cref{fig:limestone-tree}, limestone is characterized by a predominant calcite content of $> 50 \%$ and a dolomitic content of $0-50 \%$ \cite{BGS2020}. Minor constituents including quartz ($< 10 \%$), feldspar ($< 5 \%$, subdivided into alkali feldspar and plagioclase), and pyrite ($< 5 \%$) are possible.

\subsubsection{Confidence-Based Classification}
While the expert system framework effectively codifies geological knowledge through rules and mineral assemblages, real-world classifications often involve varying degrees of uncertainty. To address this, we implement a confidence-based classification mechanism that extends the base expert system by incorporating weighted confidence scores and dual thresholds for more robust decision-making. The proposed approach applies a statistical filtering mechanism after compositional weight calculation, utilizing a confidence threshold of 0.7 and a dominance threshold of 0.3 to systematically manage classification uncertainty and balance precise mineral type identification with the inherent variability of mineral compositions within geological samples.

\begin{equation}
    C_{rock} = \begin{cases}
        \text{rock type}, & \text{if } w_{max} \geq \theta_c \text{ and } (w_{max} - w_{2nd}) \geq \theta_d \\
        \text{other}, & \text{otherwise}
    \end{cases}
\end{equation}

where $w_{max}$ is the highest weight among all rock types, $w_{2nd}$ is the second highest weight, $\theta_c$ is the confidence threshold (0.7), and $\theta_d$ is the dominance threshold (0.3).

The weight calculation for each rock type incorporates the relative proportions of key minerals, as summarized in \Cref{tab:mineral_proportions}. The classification is marked as "other" when the highest confidence score falls below the confidence threshold ($\theta_c$), indicating insufficient certainty in any particular rock type, or when the difference between the highest and second highest weights is less than the dominance threshold ($\theta_d$), suggesting ambiguity between rock types. This dual-threshold approach helps prevent misclassification in cases where mineral assemblages show characteristics of multiple rock types or when the measurements do not strongly align with any single rock type's expected composition. The thresholds were empirically determined through validation with known samples to balance classification accuracy with reliability.

\begin{table}[pos=H]
\centering
\caption{Mineral Proportions for Rock Types}
\label{tab:mineral_proportions}
\begin{tabular}{@{}@{\extracolsep{\fill}}ccccc@{}}
\toprule
\textbf{Rock Type} & \textbf{Feldspars} & \textbf{Quartz} & \textbf{Micas} & \textbf{Calcite} \\
\midrule
\textbf{Granite} & 45-80\% & 20-40\% & 0-15\% & - \\
\textbf{Sandstone} & 5-25\% & >70\% & - & - \\
\multirow{2}{*}{\textbf{Limestone}} & \multirow{2}{*}{-} & \multirow{2}{*}{-} & \multirow{2}{*}{-} & >90\% (pure) \\
& & & & >50\% (dolomitic) \\
\bottomrule
\end{tabular}
\end{table}

To quantify the weight for each rock type based on a sequence of measurements, we use the following formula. For a sequence of measurements $M = \{m_1, ..., m_n\}$, the weight for each rock type $R_l$ is calculated as:
\begin{equation}
w_{R_l} = \sum_{i=1}^k \alpha_i \cdot \delta_i
\end{equation}
where:
\begin{itemize}
\item $\alpha_i$ is a weighting coefficient that represents the relative importance of the $i$-th mineral's abundance in determining the rock type classification
\item $\delta_i$ is an indicator function that equals 1 if $p_i^{min} \leq f_i(M) \leq p_i^{max}$, and 0 otherwise
\item $f_i(M)$ is the proportion of the $i$-th mineral in measurements
\item $p_i^{min}, p_i^{max}$ are the expected minimum and maximum proportions
\end{itemize}

This formula ensures that the classification process considers both the presence and the relative importance of key minerals, providing a robust framework for automated rock type identification.

\subsection{Dataset Collection and Generation}
The RRUFF database contains approximately 7,000 mineral samples, which represent 3,500 distinct mineral species. This discrepancy arises because multiple samples can belong to the same mineral species, leading to a higher number of samples than species. Our analysis revealed a significant class imbalance, with 1,522 mineral classes containing fewer than five samples and the corresponding spectral data. In this context, a `class' refers to a specific mineral species that our model aims to identify.

Given the complexity and breadth of the RRUFF dataset and the fact that no prior study has addressed this specific problem, we chose to start our investigation by focusing on specific rock types and their related minerals. This approach allows us to simplify the initial scope of the study while still addressing a meaningful and practical subset of the data.

Rather than employing standard data augmentation techniques that might introduce artifacts, we adopted a geologically-informed sampling strategy. Sampling in this study refers to the process of selecting specific mineral samples from the RRUFF database. Our sampling strategy was geologically-informed, focusing on minerals commonly found in granite, sandstone, and limestone formations (see mineral selections in \Cref{fig:granite-tree,fig:sandstone-tree,fig:limestone-tree}). This selective approach ensures that our dataset is relevant to real-world field conditions and aligns with our hybrid architecture, which integrates expert knowledge to compensate for limited training samples.

This geological context-driven selection reflects both analytical and practical considerations. The use of unoriented spectra aligns with real-world field conditions, where rock samples are not systematically oriented prior to analysis, resulting in typically random crystallographic orientations of the present mineral phases. Rather than employing a purely data-driven approach that might retain overrepresented but geologically irrelevant mineral species, our selection methodology prioritizes the minerals characteristic of these three rock types, regardless of their representation in the database. This selective sampling strategy aligns with our hybrid architecture, where expert knowledge compensates for limited training samples through geological constraints, effectively addressing both the class imbalance challenge and the need for geologically meaningful classifications.

Due to this limited sample size, we expanded our dataset using two synthetic data generation methods. First, we applied a PCA-based approach for minerals with larger datasets and second, we used a direct variation method for minerals with limited samples. Our target dataset sizes were determined by applying a 4× multiplication factor to the initial sample counts, resulting in projected totals of 439 samples for minerals associated with granite, 449 for minerals associated with limestone, 515 for minerals associated with sandstone and 123 for other minor minerals. It is important to note that these samples are mineral spectra, not rock samples. The classification of rocks is inferred from the spectral data of their constituent minerals. All spectra were obtained from processed RRUFF data, which typically includes standard preprocessing steps such as background subtraction and peak fitting, though specific processing methods may vary as the database aggregates contributions from multiple research institutions worldwide.

The effectiveness of this geological context-driven selection approach was later validated through expert-designed test cases, as detailed in \Cref{evaframework}.

\subsection{Mineral Classification}
For mineral classification, four machine learning models were implemented and tested: Support Vector Machine (SVM), Random Forest (RF), Multilayer Perceptron (MLP), and two variants of One-dimensional Convolutional Neural Networks (1D-CNN). Each model was trained using an expanded dataset of 1366 mineral samples.

Two versions of the 1D-CNN architecture were developed. The base model consists of two convolutional layers (16 and 32 channels) with ReLU activation and max pooling operations. The network processes the input spectra through these layers before passing through two fully connected layers to output predictions for the fourteen mineral classes. To handle unknown minerals, the base architecture was enhanced with an uncertainty-aware version. This model maintains the same structure but incorporates Monte Carlo dropout layers \cite{galDropoutBayesianApproximation2016} with a rate of 0.3 after each convolutional layer and the first fully connected layer. During inference, 30 stochastic forward passes are performed with enabled dropout layers, allowing the estimation of both the predictive mean and variance for uncertainty quantification. This uncertainty estimation helps identify when the model encounters mineral spectra that do not match the fourteen defined classes. Both models were trained using cross-entropy loss and the Adam optimizer with a learning rate of 0.001. To avoid overfitting, early stopping was implemented with a patience of 20 epochs, which means training was stopped if the validation loss did not improve for 20 consecutive epochs.

\subsection{Rule-Based Expert System}
The expert system was developed to automate rock classification based on Raman spectroscopy point measurements, incorporating domain knowledge from mineralogy. The system employs a set of hierarchical rules that evaluate both prediction accuracy and the presence of mineral assemblages characteristic of different rock types.

\subsubsection{Knowledge Base and Definitions}
The knowledge base $KB$ is defined as a quintuple:
\begin{equation}
    KB = (G, R, H, P, C)
\end{equation}

where:
\begin{itemize}
    \item $G = \{G_1, G_2, ..., G_K\}$ represents K distinct mineral assemblages
    \item $R = \{R_1, R_2, ..., R_L\}$ denotes L rock type classification rules
    \item $H = (V, E)$ defines the hierarchical decision tree structure
    \item $P = \{p_v | v \in V\}$ comprises classification and composition parameters
    \item $C = \{C_1, C_2, ..., C_J\}$ represents both compositional and confidence constraints
\end{itemize}

For any mineral $m$ and group $G_i$ with weight $w_i$, the weighted membership function is defined as:
\begin{equation}
\mu_{G_i}(m, w_i) = 
\begin{cases}
    w_i & \text{if } m \in G_i \text{ and } p_i^{min} \leq f_i(m) \leq p_i^{max} \\
    0 & \text{otherwise} 
\end{cases}
\end{equation}

\subsubsection{Compositional Rules and Constraints}
The confidence-based compositional requirements for each rock type are formalized through constraint functions:
\begin{equation}
    C_j(M, G_i, w_{R_l}) = f_j(n_i, \alpha_j, w_{R_l}) \geq \beta_j
\end{equation}
where 
\begin{itemize}
    \item $f_j$ represents a constraint function
    \item $n_i$ is the count of minerals from group $G_i$ in measurements $M$
    \item $\alpha_j$ is a parameter vector
    \item $w_{R_l}$ is the importance weight for rule $R_l$
    \item $\beta_j$ defines the threshold value
\end{itemize}

The classification rule incorporating confidence thresholds is expressed as:
\begin{equation}
    R_l(M) = \left(\bigwedge_{j \in \mathcal{J}_l} C_j(M, G_{i_j}, w_{i_j})\right) \land C_{conf}(w_{max}, w_{2nd})
\end{equation}
$
\text{where } C_{conf}(w_{max}, w_{2nd}) = (w_{max} \geq \theta_c) \land (w_{max} - w_{2nd} \geq \theta_d)
$

\subsubsection{Mineral Assemblages}
Mineral assemblages for each rock type are represented as weighted sets with composition ranges:
\begin{equation}
    A_{R_l} = \{(G_i, w_i, [p_i^{min}, p_i^{max}]) | G_i \in G, w_i \in [0,1], p_i^{min}, p_i^{max} \in [0,1]\}
\end{equation}
where $w_i$ represents the diagnostic weight and $[p_i^{min}, p_i^{max}]$ defines the valid composition range of mineral group $G_i$.

\subsubsection{Weighted Importance Model}
Given the weighted mineral assemblages $A_{R_l}$, we define the confidence-adjusted probability of a rock type rule $R_l$ given measurements $M$ as:
\begin{equation}
P(R_l|M) = \prod_{(G_i, w_i, [p_i^{min}, p_i^{max}]) \in A_{R_l}} w_i^{n_i} \cdot \delta_i
\end{equation}
where:
\begin{itemize}
\item $w_i$ is derived from geological composition ranges and importance weights
\item $n_i = \text{count}(M, G_i)$ is the number of minerals from group $G_i$ in measurements $M$
\item $\delta_i$ is an indicator function that equals 1 if $p_i^{min} \leq f_i(M) \leq p_i^{max}$, and 0 otherwise
\end{itemize}

\subsubsection{Evaluation Framework} \label{evaframework}
To evaluate the expert system's performance under the constraints of open-source geological datasets, 10 test cases for each rock type were utilized, specifically designed by an expert geologist. These cases represent both confident and non-confident classification scenarios that occur in real geological settings. For confident cases, mineral assemblages that unambiguously indicate specific rock types were selected, representing typical compositions found in well-documented geological formations. In contrast, non-confident cases were designed with mineral assemblages that clearly indicate different rock types, challenging the system's classification capabilities. Confusion matrix analysis was used to quantitatively assess classification performance. This analysis measures true positives (correct rock type identification), false positives (incorrect identification of rock type), true negatives (correct rejection of wrong rock type), and false negatives (incorrect rejection of correct rock type). From these measurements, standard performance metrics, including accuracy, precision, recall, and F1-score, were calculated to provide a comprehensive assessment of the system's classification capabilities.

\section{Results \& Discussion}\label{results}
The experimental evaluation of our mineral assemblage-based classification framework encompasses three key aspects: the performance of individual mineral identification, the efficacy of the integrated rock classification system and the analysis of current limitations. Quantitative results from both the baseline and uncertainty-aware models are presented, followed by a comparative analysis of their performance across a validation set.

As highlighted by Resch et al. \cite{reschTunnelExcavationMaterial2009}, excavated materials from tunnel projects have diverse reuse pathways: limestone can serve as raw material for the steel industry and feedstuffs production, soil can be utilized for brick manufacturing, rock dust has applications in agricultural land improvement, and certain rock types yield valuable industrial minerals such as mica for the paint industry. Therefore, accurate classification of rock types is crucial for optimizing the reuse of excavated materials and minimizing environmental impact.

\subsection{Mineral classification}
\begin{figure}[pos=H]
    \centering
    \includegraphics[width=0.8\linewidth]{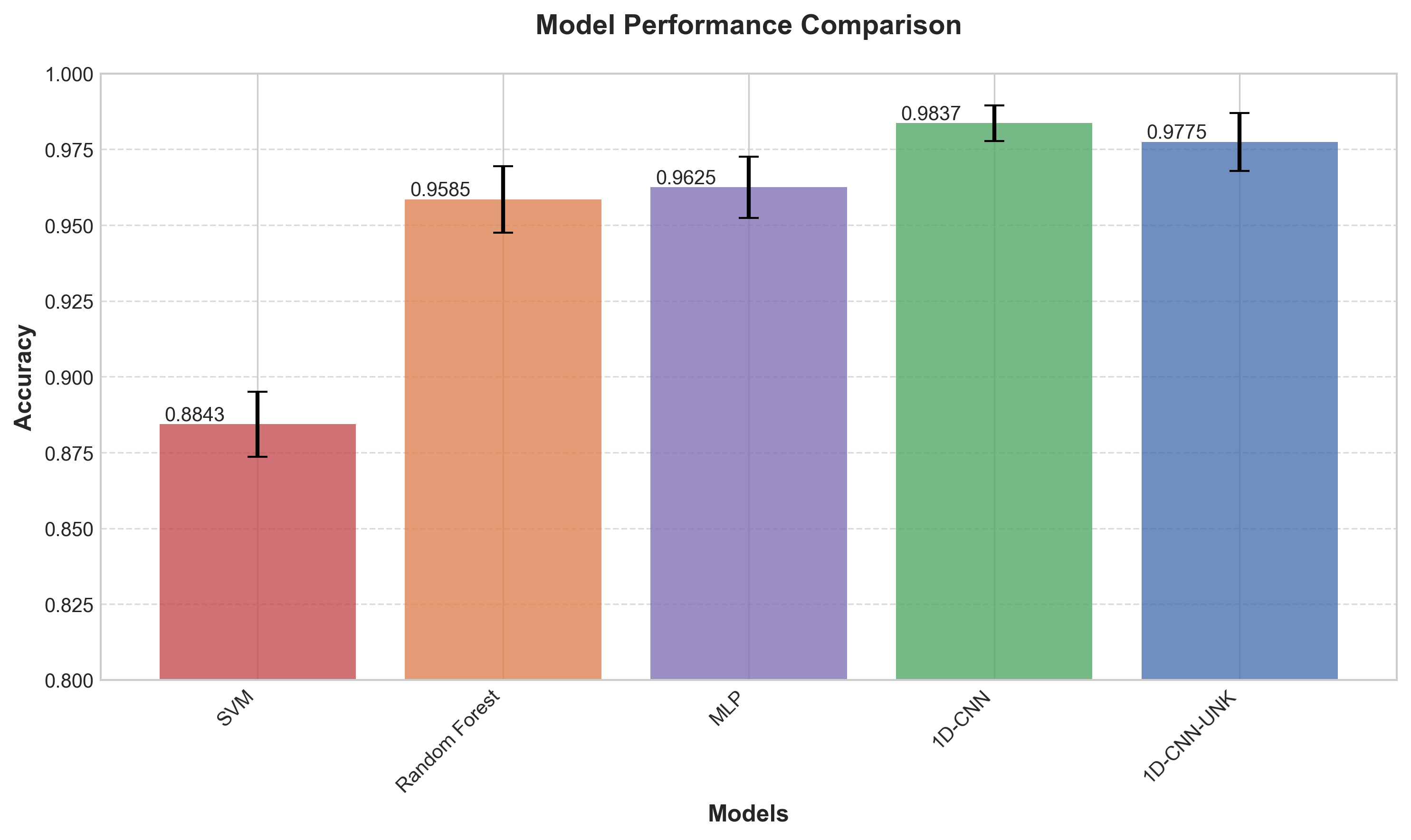}
    \caption{Performance comparison of different machine learning models for minerals classification. The bar plot shows the mean accuracy with standard error bars for five different classifiers: Support Vector Machine (SVM), Random Forest, Multilayer Perceptron (MLP), One-dimensional Convolutional Neural Network (1D-CNN), and 1D-CNN with unknown class handling (1D-CNN-UNK). Results are averaged across five-fold cross-validation.}
    \label{fig:models}
\end{figure}

The performance of different machine learning models for mineral classification was first evaluated. \Cref{fig:models} shows the mean accuracy across five-fold cross-validation for SVM, Random Forest, MLP, 1D-CNN, and 1D-CNN-UNK models. The 1D-CNN model achieved the highest accuracy at 98.37\%, while the 1D-CNN-UNK model showed slightly lower performance at 97.75\%. Both models outperformed the baseline approaches (SVM at 88.43\%, Random Forest at 95.85\%, and MLP at 96.25\%). Although a confusion matrix would provide detailed per-mineral performance, overall accuracy was focused on the main metric, since the performance of the integrated system is the primary concern.

\subsection{Integrated System Performance}

\begin{figure}[pos=H]
    \centering
    \begin{subfigure}[b]{0.48\linewidth}
        \centering
        \includegraphics[width=\linewidth]{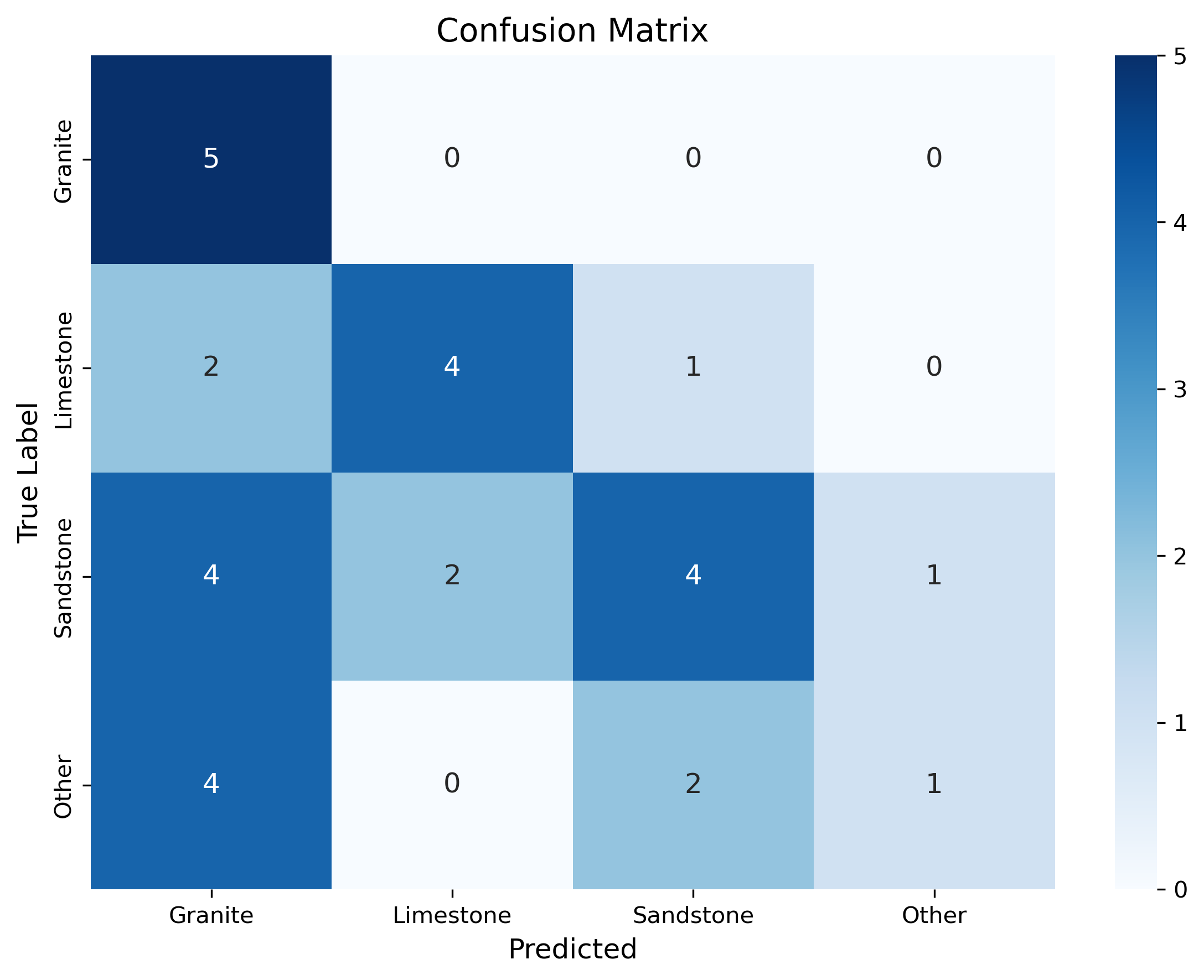}
        \caption{knowledge-guided 1D-CNN}
        \label{fig:confusion_baseline}
    \end{subfigure}
    \hfill
    \begin{subfigure}[b]{0.48\linewidth}
        \centering
        \includegraphics[width=\linewidth]{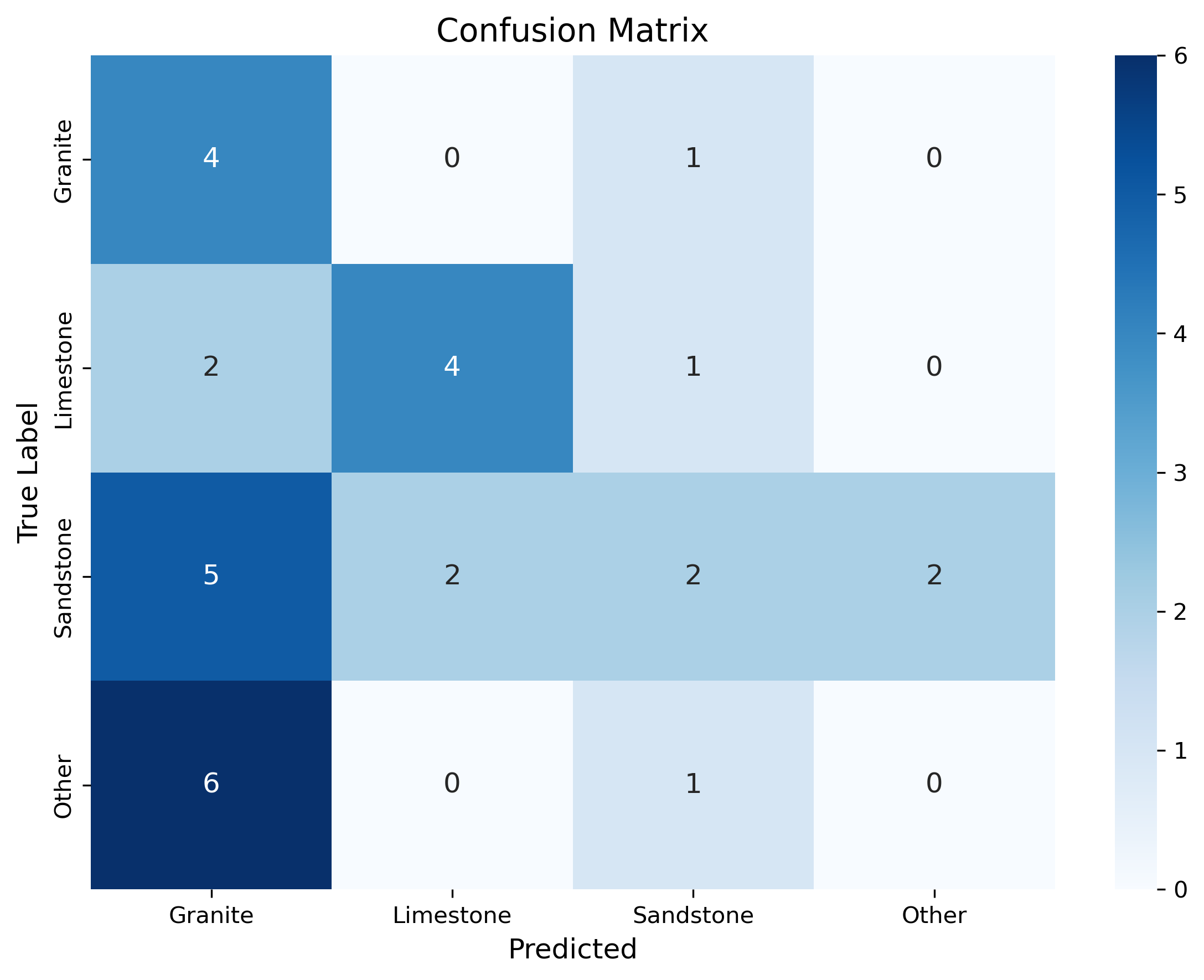}
        \caption{uncertainty-aware knowledge-guided 1D-CNN}
        \label{fig:confusion_uncertainty}
    \end{subfigure}
    \caption{Confusion matrices comparing rock type classification between (a) knowledge-guided 1D-CNN and (b) uncertainty-aware knowledge-guided 1D-CNN. Darker colors indicate higher numbers of samples.}
    \label{fig:confusion_comparison}
\end{figure}

The comparative analysis of the baseline model and the uncertainty-aware 1D-CNN model reveals distinct performance patterns across rock classifications (n=30 samples). Both architectures demonstrate varying classification efficacy across different lithologies, with notable differences in their respective error distributions and classification confidence patterns. The baseline model (\Cref{fig:confusion_baseline}) achieves 33.3\% precision and 100\% recall for granite identification (n=5), while the uncertainty-aware variant (\Cref{fig:confusion_uncertainty}) exhibits 23.5\% precision and 80\% recall. For limestone samples (n=7), both models maintain consistent performance metrics (precision: 66.7\%, recall: 57.1\%, F1-score: 0.62), correctly classifying four samples. The most substantial performance differential manifests in sandstone classification, where the uncertainty-aware model demonstrates reduced classification accuracy (precision: 40\%, recall: 18.2\%, F1-score: 0.25) compared to the baseline architecture (precision: 57.1\%, recall: 36.4\%, F1-score: 0.44).

Quantitative analysis of misclassification patterns reveals systematic variations between the two models. The uncertainty-aware variant exhibits increased granite misclassifications, particularly for samples from the "Other" category (n=6 versus n=4 in baseline) and sandstone samples (n=5 versus n=4 in baseline). This systematic misclassification pattern between sandstone and granite samples indicates underlying limitations in spectral feature discrimination. These experimental results reveal a fundamental challenge: the inherent mismatch between training data composition (single mineral spectra) and the target classification objective (whole rock assemblages). This disparity manifests itself in three critical aspects. First, the uncertainty-aware architecture demonstrates modified classification boundaries for sandstone identification, resulting in increased misclassifications toward granite categories ($\Delta n= +1$ sample). Second, while the 1D-CNN architecture effectively learns individual mineral spectral signatures, the classification accuracy decreases when processing complex mineral assemblages within whole rock samples ($\Delta F1_{sandstone} = -0.19$). Third, the expert system rules, despite incorporating established geological constraints, demonstrate limited effectiveness in differentiating between compositionally similar rock types (P(granite) < 35\% for both models).

The observed performance patterns manifested most significantly in cases where rocks share similar mineral constituents in varying proportions, such as granite and sandstone. The uncertainty-aware variant's performance metrics indicate that the primary challenge extends beyond the disparity between single-mineral training data and whole-rock classification objectives. Future research priorities should focus on the development of comprehensive mineral assemblage training datasets that capture spectral interactions within whole rock samples. Additionally, measuring the relative amounts of different minerals could enhance the analysis, providing more nuanced insights into rock composition and potentially improving classification accuracy.

These findings indicate that improving classification accuracy requires addressing both fundamental data representation challenges and the methodology for handling compositional uncertainty in rock sample analysis.

\subsection{Limitation}
Although the method effectively detects the presence of specific minerals through their characteristic Raman spectral signatures, it faces several key challenges: (i) compositional ambiguity and (ii) classification constraints.

Because different rock types can share similar mineral assemblages while having distinct geological origins and classifications, a classification overlap was observed in the results. As demonstrated by our confusion matrices (\Cref{fig:confusion_comparison}), both granites and sandstones exhibit significant classification overlap due to their shared mineral constituents, despite different proportions. The binary (presence/absence) nature of the current approach does not capture the subtle variations in mineral proportions that often distinguish different rock types. This limitation is particularly evident in the low precision rates for granite classification (<35\%) and the systematic patterns of misclassifications observed between compositionally similar rock types.

\section{Conclusion}\label{conclusion}
This investigation advances automated geological classification through a novel integration of Raman spectroscopy and knowledge-enhanced deep learning methodologies. Our quantitative analysis, based on a limited dataset of 30 samples, demonstrates the effectiveness of the hybrid mineral-to-rock classification framework. The 1D-CNN architecture achieved 98.37±0.006\% accuracy in mineral identification, while the uncertainty-aware variant achieved 97.75±0.010\% accuracy. The implementation of confidence thresholds in the knowledge system, governed by the weighting function introduced in this paper, provides systematic discrimination between compositionally similar rock types. This is particularly evident in the classification of limestone samples, with a precision of 66.7\%, recall of 57.1\%, and an F1-score of 0.62.

The methodological framework addresses some of the fundamental challenges in automated rock classification through the systematic integration of spectroscopic data analysis and expert geological knowledge. The achieved accuracy demonstrates the effectiveness of our knowledge-enhanced approach in compensating for data sparsity through expert rule integration. However, it is important to note that the small sample size (n=30) limits the generalizability of these findings, and further validation with a larger dataset is necessary.

While this preliminary investigation establishes methodological feasibility, it also highlights clear pathways for future development. The transition from controlled laboratory conditions to conveyor belt operations presents opportunities for technological advancement. Integrating multiple sensing modalities and optimized data acquisition protocols could reduce the amount of laboratory experiments required, although laboratory validation will remain essential. These developments, coupled with our demonstrated success in mineral identification and classification, provide a robust framework for advancing automated geological characterization systems.

This study serves as a foundational effort in the subfield of petrology, where open datasets are currently limited. By demonstrating the potential of our approach, we aim to inspire the creation of comprehensive, open-source mineral assemblage datasets. Such datasets would enable more robust validation of automated classification systems and further enhance the advantages of our knowledge-enhanced approach. Additionally, incorporating the relative ratios of minerals in the analysis could improve classification accuracy and address the compositional ambiguity observed in this study.

The established methodology not only addresses current industrial needs but also lays the groundwork for more comprehensive rock type classification systems, positioning this research at the forefront of automated geological analysis.

\clearpage
\appendix
\renewcommand{\theequation}{\Alph{section}.\arabic{equation}} 
\renewcommand{\thefigure}{\Alph{section}.\arabic{figure}}     
\renewcommand{\thetable}{\Alph{section}.\arabic{table}}       
\setcounter{equation}{0}  
\setcounter{figure}{0}    
\setcounter{table}{0}     
\renewcommand{\thesection}{\appendixname~\Alph{section}} 
\section{Rock Composition}

    \begin{table}[pos=H]
        \centering
        \onehalfspacing
        \setlength\tabcolsep{3pt}
        \rotatebox{90}{%
        \begin{minipage}{\linewidth}
        \caption{Suggested rock composition of 30 samples.}
        \resizebox{1.0\linewidth}{!}{%
        \begin{tabular}{@{}|l|l|l|l|l|l|l|l|l|l|l|l|p{4cm}|@{}}
            \hline
            \textbf{No} & \textbf{1}  & \textbf{2}  & \textbf{3} & \textbf{4} & \textbf{5}    & \textbf{6}  & \textbf{7} & \textbf{8} & \textbf{9}  & \textbf{10} & \textbf{Result} & \textbf{Reasons}                                                                                           \\ \hline
            Example 1   & Albite      & Anorthite   & Quartz     & Quartz     & Annite        & Muscovite   & Quartz     & Albite     & Annite      & Orthoclase  & Granite         &                                                                                                            \\ \hline
            Example 2   & Albite      & Quartz      & Annite     & Annite     & Muscovite     & Orthoclase  & Orthoclase & Quartz     & Quartz      & Anorthite   & Granite         &                                                                                                            \\ \hline
            Example 3   & Jadeite     & Quartz      & Quartz     & Jadeite    & Orthoclase    & Jadeite     & Anorthite  & Quartz     & Annite      & Quartz      & Not a Granite   & because of the jadeite which forms in metamorphic rocks                                                    \\ \hline
            Example 4   & Orthoclase  & Muscovite   & Annite     & Orthoclas  & Albite        & Anorthite   & Muscovite  & Annite     & Quartz      & Phlogopite  & Granite         &                                                                                                            \\ \hline
            Example 5   & Annite      & Phlogopite  & Omphacite  & Albite     & Omphacite     & Quartz      & Annite     & Omphacite  & Omphacite   & Omphacite   & Not a Granite   & because of the omphacite which forms in metamorphic rocks                                                  \\ \hline
            Example 6   & Annite      & Quartz      & Quartz     & Annite     & Albite        & Anorthite   & Orthoclase & Albite     & Annite      & Muscovite   & Granite         &                                                                                                            \\ \hline
            Example 7   & Quartz      & Annite      & Muscovite  & Quartz     & Anorthite     & Orthoclase  & Quartz     & Albite     & Quartz      & Annite      & Granite         &                                                                                                            \\ \hline
            Example 8   & Orthoclase  & Glaucophane & Anorthite  & Albite     & Glaucophane   & Phlogopite  & Quartz     & Quartz     & Glaucophane & Glaucophane & Not a Granite   & because of the glaucophane which forms in metamorphic rocks                                                \\ \hline
            Example 9   & Anorthite   & Quartz      & Staurolite & Almandine  & Orthoclase    & Phlogopite  & Muscovite  & Albite     & Quartz      & Annite      & Not a Granite   & because of staurolite and almandine, both form in metamorphic rocks                                        \\ \hline
            Example 10  & Almandine   & Annite      & Quartz     & Anorthite  & Albite        & Orthoclase  & Muscovite  & Garnet     & Quartz      & Annite      & Not a Granite   & because of Garnet                                                                                          \\ \hline
            Example 11  & Quartz      & Calcite     & Ortholcase & Albite     & Anorthite     & Pyrite      & Annite     & Phlogopite & Muscovite   & Tourmaline  & Sandstone       &                                                                                                            \\ \hline
            Example 12  & Albite      & Orthoclase  & Anorthite  & Tourmaline & Rutile        & Muscovite   & Quartz     & Calcite    & Annite      & Anorthite   & Sandstone       &                                                                                                            \\ \hline
            Example 13  & Sanidine    & Albite      & Sanidine   & Quartz     & Calcite       & Quartz      & Albite     & Anorthite  & Jadeite     & Annite      & Not a Sandstone & because of sanidine and jadeite, sanidine is in magmatic rocks and jadeite is in metamorphic rocks         \\ \hline
            Example 14  & Pyrite      & Orthoclase  & Quartz     & Calcite    & Quartz        & Albite      & Anorthite  & Tourmaline & Rutile      & Annite      & Sandstone       &                                                                                                            \\ \hline
            Example 15  & Albite      & Anorthite   & Anorthite  & Phlogopite & Muscovite     & Annite      & Quartz     & Orthoclase & Quartz      & Quartz      & Sandstone       &                                                                                                            \\ \hline
            Example 16  & Orthoclase  & Albite      & Anorthite  & Annite     & Muscovite     & Quartz      & Quartz     & Quartz     & Tourmaline  & Annite      & Sandstone       &                                                                                                            \\ \hline
            Example 17  & Quartz      & Anorthite   & Orthoclase & Orthoclase & Phlogopite    & Pyrope      & Quartz     & Omphacite  & Quartz      & Annite      & Not a Sandstone & because of pyrope and omphacite, both minerals form in metamorphic rocks                                   \\ \hline
            Example 18  & Albite      & Phlogopite  & Orthoclase & Albite     & Muscovite     & Annite      & Calcite    & Quartz     & Albite      & Pyrite      & Sandstone       &                                                                                                            \\ \hline
            Example 19  & Glaucophane & Albite      & Quartz     & Quartz     & Albite        & Glaucophane & Muscovite  & Quartz     & Calcite     & Annite      & Not a Sandstone & because of glaucophane which forms in metamorphic rocks                                                    \\ \hline
            Example 20  & Quartz      & Muscovite   & Tourmaline & Quartz     & Calcite       & Albite      & Tourmaline & Annite     & Muscovite   & Orthoclase  & Sandstone       &                                                                                                            \\ \hline
            Example 21  & Calcite     & Quartz      & Dolomite   & Calcite    & Calcite       & Anorthite   & Albite     & Calcite    & Calcite     & Dolomite    & Limestone       &                                                                                                            \\ \hline
            Example 22  & Andalusite  & Calcite     & Calcite    & Albite     & Orthoclase    & Calcite     & Calcite    & Kyanite    & Pyrite      & Calcite     & Not a Limestone & because of andalusite and kyanite which both form in metamorphic rocks                                     \\ \hline
            Example 23  & Quartz      & Calcite     & Calcite    & Calcite    & Calcite       & Calcite     & Calcite    & Quartz     & Calcite     & Calcite     & Limestone       &                                                                                                            \\ \hline
            Example 24  & Dolomite    & Calcite     & Calcite    & Epidote    & Rhodochrosite & Calcite     & Calcite    & Calcite    & Calcite     & Calcite     & Not a Limestone & because of the presence of epidote it is more likely a marble, epidote and marble are in metamorphic rocks \\ \hline
            Example 25  & Calcite     & Calcite     & Calcite    & Calcite    & Calcite       & Calcite     & Calcite    & Calcite    & Calcite     & Calcite     & Limestone       &                                                                                                            \\ \hline
            Example 26  & Quartz      & Calcite     & Quartz     & Albite     & Calcite       & Pyrite      & Quartz     & Calcite    & Calcite     & Dolomite    & Limestone       &                                                                                                            \\ \hline
            Example 27  & Calcite     & Quartz      & Dolomite   & Calcite    & Calcite       & Calcite     & Dolomite   & Quartz     & Dolomite    & Quartz      & Limestone       &                                                                                                            \\ \hline
            Example 28  & Sanidine    & Quartz      & Sanidine   & Calcite    & Dolomite      & Sanidine    & Calcite    & Calcite    & Calcite     & Calcite     & Not a Limestone & because of sanidine which is in magmatic rocks                                                             \\ \hline
            Example 29  & Calcite     & Calcite     & Dolomite   & Albite     & Dolomite      & Calcite     & Calcite    & Calcite    & Dolomite    & Calcite     & Limestone       &                                                                                                            \\ \hline
            Example 30  & Calcite     & Calcite     & Dolomite   & Calcite    & Calcite       & Calcite     & Dolomite   & Calcite    & Calcite     & Calcite     & Limestone       &                                                                                                            \\ \hline
        \end{tabular}%
        }
    \end{minipage}
        }
    \end{table}

\clearpage
\section{Notations}

\setcounter{equation}{0}  
\setcounter{figure}{0}    
\setcounter{table}{0}     

\begin{table}[pos=H]
    \caption{Table of Notation}
    \label{tab:notation}
    \centering
    \singlespacing
    \begin{tabular}{p{0.15\textwidth} p{0.75\textwidth}}
    \hline
    \textbf{Symbol} & \textbf{Description} \\
    \hline
    \multicolumn{2}{l}{\textit{Measurements and Classifications}} \\
    $M$ & Sequence of measurements $\{m_1, ..., m_n\}$ \\
    $m_i$ & Individual measurement in sequence \\
    $f_i(M)$ & Proportion of the $i$-th mineral in measurements \\
    $\delta_i$ & Indicator function for mineral proportion constraints \\
    $w_{R_l}$ & Classification weight for rock type rule $R_l$ \\
    $w_{max}$ & Maximum classification weight \\
    $w_{2nd}$ & Second-highest classification weight \\
    \hline
    \multicolumn{2}{l}{\textit{Mineral Assemblages}} \\
    $A_{R_l}$ & Weighted mineral assemblage for rule $R_l$ \\
    $n_i$ & Number of minerals from group $G_i$ in measurements \\
    $p_i^{min}$ & Minimum expected proportion for $i$-th mineral \\
    $p_i^{max}$ & Maximum expected proportion for $i$-th mineral \\
    \hline
    \multicolumn{2}{l}{\textit{Knowledge Base Components}} \\
    $KB$ & Knowledge base \\
    $G$ & Set of mineral assemblages \\
    $R$ & Set of rock type classification rules \\
    $R_l$ & Individual classification rule, where $l \in \{1,...,L\}$ \\
    $H$ & Hierarchical decision tree structure \\
    $P$ & Set of classification and composition parameters \\
    $C$ & Set of compositional and confidence constraints \\
    $V, E$ & Vertices and edges in decision tree \\
    \hline
    \multicolumn{2}{l}{\textit{Parameters}} \\
    $\alpha_i$ & Weighting coefficient representing relative importance of $i$-th mineral's abundance \\
    $\theta_c$ & Confidence threshold \\
    $\theta_d$ & Separation threshold \\
    $K, L, J$ & Number of assemblages, rules, and constraints \\
    \hline
    \multicolumn{2}{l}{\textit{Mathematical Operators}} \\
    $\prod$ & Product operator over a sequence of terms \\
    $\bigwedge$ & Logical AND operator over a sequence of conditions \\
    \hline
    \end{tabular}
    \end{table}
    
    \newpage

\begin{table}[pos=H]
    \caption{Mapping of Rock Classification Trees to Mathematical Framework}
    \label{tab:rock_mapping}
    \centering
    \singlespacing
    \begin{tabular}{p{0.12\textwidth} p{0.2\textwidth} p{0.16\textwidth} p{0.16\textwidth} p{0.16\textwidth}}
    \hline
    \textbf{Component} & \textbf{Mathematical Form} & \textbf{Granite} & \textbf{Sandstone} & \textbf{Limestone} \\
    \hline
    \multicolumn{5}{l}{\textit{Mineral Groups ($G_i \in G$)}} \\
    Primary Groups & $G_1, G_2, ...$ & Quartz, Feldspars, Micas & Quartz, Feldspars, Calcite & Calcite, Dolomite \\
    \hline
    \multicolumn{5}{l}{\textit{Weighted Assemblages ($A_{R_l}$)}} \\
    Major & $(G_i, w_{R_l}, [p_i^{min}, p_i^{max}])$ & (Quartz, 0.6, [0.20, 0.40]) & (Quartz, 0.9, [0.70, 1.00]) & (Calcite, 0.9, [0.90, 1.00]) \\
    Essential & $(G_i, w_{R_l}, [p_i^{min}, p_i^{max}])$ & (Feldspars, 0.8, [0.45, 0.80]) & (Feldspars, 0.5, [0.05, 0.25]) & (Dolomite, 0.7, [0.10, 0.50]) \\
    Accessory & $(G_i, w_{R_l}, [p_i^{min}, p_i^{max}])$ & (Micas, 0.3, [0, 0.15]) & (Mica, 0.2, [0.02, 0.03]) & (Quartz, 0.2, [0, 0.10]) \\
    \hline
    \multicolumn{5}{l}{\textit{Classification Rules ($R_l \in R$)}} \\
    Case Rules & $R_l(M)$ & "Granite if Feldspars 45-80\%" & "Sandstone if Quartz >70\%" & "Pure Limestone if Calcite >90\%" \\
    Constraints & $C_j(M, G_i, w_{R_l})$ & "AND Quartz 20-40\%" & "AND Feldspars 5-25\%" & "OR Dolomitic if Dolomite 10-50\%" \\
    \hline
    \multicolumn{5}{l}{\textit{Hierarchical Structure ($H = (V, E)$)}} \\
    Parent-Child & $(v_p, v_c) \in E$ & Feldspars → Orthoclase/Albite & Feldspars → Orthoclase/Albite & Limestone → Pure/Dolomitic \\
    Leaf Nodes & $v \in V$ & Orthoclase, Albite, Anorthite & Orthoclase, Albite, Anorthite & Orthoclase, Albite, Anorthite \\
    \hline
    \end{tabular}
    \end{table}

\clearpage
\printcredits

\section*{Code Availability}
The code developed in this research is available on GitHub: \url{https://github.com/iyeszin/rocks_evaluation.git}. This repository contains all evaluation rock data and codes required to reproduce the results presented in this paper.

\section*{Declaration of competing interest}
The authors declare that they have no known competing financial
interests or personal relationships that could have appeared to influence the work reported in this paper.

\section*{Acknowledgments}
We gratefully acknowledge support for the work in this paper from Austrian Federal Ministry for Climate Action, Environment, Energy, Mobility, Innovation and Technology (BMK) and Austrian Research Promotion Agency (FFG project number: FO999907606) as part of the FTI Circular Economy Initiative (3rd call for proposals, 2023) within open4innovation.

\section*{Data availability}
Data is available from the RRUFF Project website: \url{https://rruff.info/zipped_data_files/}. We used high-quality unoriented Raman spectra from RRUFF. 

\section*{Declaration of Generative AI and AI-assisted technologies in the writing process}
\label{sec:ai-declaration}

During the preparation of this work, the authors used Anthropic's Claude to assist with editing the manuscript and improving its readability and style. The authors then carefully reviewed, revised, and edited the content as needed, taking full responsibility for the final version of the publication.

\bibliographystyle{elsarticle-num-names} 

\bibliography{literature}

\end{document}